\documentclass[a4paper,twoside]{article}

\usepackage{epsfig}
\usepackage{subcaption}
\usepackage{calc}
\usepackage{ulem}
\usepackage{amsmath, amssymb, amsfonts}
\usepackage{tikz}
\usetikzlibrary{automata, positioning, arrows,calc, decorations.pathreplacing}
\usepackage{natbib}
\usepackage{amsthm}
\newtheorem{definition}{Definition}

\newtheorem{example}{Example}
\newtheorem{remark}{Remark}

\newcommand{\A}{\mathcal{A}}
\newcommand{\Sy}{\mathcal{S}}
\newcommand{\G}{\mathcal{G}}
\newcommand{\FO}{FO(\mathcal{V}, =, P_\G)}
\usepackage[colorlinks=true, linkcolor=blue, citecolor=blue, filecolor=blue, urlcolor=blue]{hyperref}
\usepackage{amssymb}
\usepackage{amstext}
\usepackage{amsmath}
\usepackage{amsthm}
\usepackage{multicol}
\usepackage{pslatex}
\usepackage{apalike}
\usepackage{algorithm2e}
\usepackage[bottom]{footmisc}
\renewcommand{\emph}[1]{\textit{#1}}
\usepackage{array}
\usepackage{graphicx}

\usepackage{lastpage}
\usepackage{fancyhdr}

\usepackage{tabularx}
\usepackage{array}

\tikzset{
    >=stealth,                  
    every picture/.style={scale=0.6, xshift=-1cm}, 
    every node/.style={scale=0.8}, 
    initial text=,              
    node distance=1.6cm,        
    highlight/.style={line width=0.6mm, draw=brown, text=brown},
    intheory/.style={draw=blue, text=blue},
    invented/.style={draw=brown, text=brown}
}

\usepackage{SCITEPRESS}     

\begin{document}

\title{An Automata-Based Method to Formalize Psychological Theories : \\
The Case Study of Lazarus and Folkman’s Stress Theory} 

\author{\authorname{Alain Finkel\sup{1}\orcidAuthor{0000-0003-2482-6141}, Gaspard Fougea\sup{1}\orcidAuthor{0009-0004-8357-5340} and Stéphane Le Roux\sup{1}}
\affiliation{\sup{1} Université Paris-Saclay, CNRS, ENS Paris-Saclay, LMF, 91190, Gif-sur-Yvette, France}
\email{\{gaspard.fougea, alain.finkel, stephane.le\_roux\}@ens-paris-saclay.fr}
}


\abstract{Formal models are important for theory-building, enhancing the precision of predictions and promoting collaboration. Researchers have argued that there is a lack of formal models in psychology. We present an automata-based method to formalize psychological theories, i.e. to transform verbal theories into formal models. This approach leverages the tools of theoretical computer science for formal theory development, for verification, comparison, collaboration, and modularity. We exemplify our method on Lazarus and Folkman's theory of stress, showcasing a step-by-step modeling of the theory. }

\onecolumn \maketitle \normalsize \setcounter{footnote}{0} \vfill

\section{INTRODUCTION}
\label{sec:introduction}

\paragraph{Context:}
For decades, some researchers have been arguing that the field of psychology is in a state of crisis. \citep{meehl_theoretical_nodate} expressed that "Theories in “soft” areas of psychology lack the cumulative character of scientific knowledge". The study \citep{open_science_collaboration_estimating_2015} showed that about half of the literature in psychology does not replicate. Several researchers express that these crises stem partly from a lack of formal models in psychology \citep{meehl_theoretical_nodate, borsboom_theory_2021, robinaugh2021invisible}. 

Most psychological theories are \emph{verbal} theories, i.e. expressed in natural language, with all of its imprecision \citep{haslbeck2022modeling}. This imprecision makes it difficult to make precise predictions \citep{robinaugh2021invisible}, which are necessary to validate or falsify a theory. Furthermore, according to \citep{robinaugh2021invisible}, \emph{"verbal theories do not lend themselves to collaborative development"}, explaining why theories in psychology are like "toothbrushes" (“no self-respecting person wants to use anyone else’s”, \citep{mischel_toothbrush_2008}). 

\citep{smaldino_better_2019}, \citep{muthukrishna_problem_2019}) and \citep{robinaugh2021invisible} argue that formal models will address some of the theoretical issues of psychology: \emph{"We argue that formal theories provide this much needed set of tools, equipping researchers with tools for thinking, evaluating explanation, enhancing measurement, informing theory development, and promoting the collaborative construction of psychological theories"} \citep{robinaugh2021invisible}.

\paragraph{Expected properties of a formal model for psychology:} Following \citep{wing2006computational}, we are inspired by the \emph{computational thinking} that involves decomposing complex theories, identifying patterns, focusing on essential details, designing step-by-step solutions and evaluating their effectiveness. 
Let us propose a list of desirable properties that, we believe, formal method for psychological theories should satisfy.
\begin{enumerate}
    \item Openness to all psychological theories, both cognitive and behavioral,
    \item Modularity (easy to modify, compose, and refine),
    \item Having a formal semantics,
    \item Formal composition and refinement,
    \item Capability to handle large systems,
    \item Possibility of step-by-step simulation,
    \item Formal verification of properties (psychological model checking) with the use of automatic tools,
    \item Formal (and automatic) comparison of models, with automatic determination of compatibility between theories.
\end{enumerate}



\paragraph{Properties satisfied by psychological models:} We identify 13 well-known frameworks used in psychology which we will list below.
For each of these frameworks, we examine which of the eight above-mentioned properties are satisfied.

\begin{figure*}
\noindent
\begin{center}

\resizebox{0.95\textwidth}{!}{%
\begin{tabular}{|>{\raggedright\arraybackslash}p{3.5cm}|c|c|c|c|c|c|c|c|}
\hline
\textbf{Model} & \textbf{1} & \textbf{2} & \textbf{3} & \textbf{4} & \textbf{5} & \textbf{6} & \textbf{7} & \textbf{8} \\
\hline
ACT-R & Yes & Yes & Partially & Partially & Partially & Yes & Partially & Partially \\
\hline
Friston's Free Energy Principle & Yes & Partially & Yes & Partially & Yes & Partially & No & No \\
\hline
GWT (Global Workspace Theory) & Partially & Partially & No & No & Partially & Partially & No & No \\
\hline
LIDA & Yes & Yes & Partially & Partially & Yes & Yes & No & No \\
\hline
Working Memory Model & No & Yes & No & No & Partially & No & No & No \\
\hline
Dual-Process Theory & Partially & No & No & No & No & No & No & No \\
\hline
Bayesian Networks & Yes & Yes & Yes & Yes & Yes & Yes & Partially & Yes \\
\hline
Neural Networks & Yes & Partially & No & Partially & Yes & No & No & No \\
\hline
Dynamic Systems & Yes & Partially & Yes & Partially & Yes & Yes & Partially & No \\
\hline
Game Theory & Yes & Yes & Yes & Yes & Yes & Yes & Partially & Yes \\
\hline
Propositional/Modal Logic & Partially & Yes & Yes & Yes & Partially & Yes & Yes & Yes \\
\hline
Information Processing Models & Partially & Yes & Yes & Yes & Partially & Yes & Partially & Yes \\
\hline
MDPs & Yes & Yes & Yes & Yes & Partially & Yes & Partially & Yes \\
\hline
\end{tabular}
} 
\end{center}
    \caption{Models and properties}
    \label{fig:my_label}
\end{figure*}
%
Let us recall briefly these 13 models.

\noindent
ACT-R (Adaptive Control of Thought - Rational) simulates human cognitive processes \citep{Anderson2004} with formal and non formal models. ACT-R satisfies modularity, step-by-step simulation, and openness to theories but lacks formalization, verification, and automatic comparison.
%
%
Friston's Predictive Coding and Free Energy Principle postulates that the brain minimizes prediction errors \citep{Friston2009}, and it is used to explain perception and decision-making. Friston's Free Energy Principle excels in formal aspects and adaptability to complex systems but struggles with modularity and automatic verification.
%
%
GWT (Global Workspace Theory), initiated by Baars, suggests consciousness arises when information is shared across the brain \citep{Baars1997,Baars2005}. 
LIDA (Learning Intelligent Distribution Agent) is based on global workspace theory, simulating attention, memory, and decision-making \citep{Franklin2007}. LIDA is modular and handles large systems but lacks formal semantics and verification tools. 
%
The Working Memory Model breaks memory into components for processing \citep{Baddeley1974}. Global Workspace Theory is conceptual, suited for theoretical exploration but lacks formalization and automatic verification.
Dual-Process Theory distinguishes between fast, automatic thinking (System 1) and slow, deliberate thinking (System 2) \citep{Kahneman2011}. Dual-Process Theory is not formalized, remaining a conceptual framework. 
Bayesian networks are widely used to model decision-making, probabilistic reasoning, and learning \citep{Pearl2018}. Bayesian networks satisfy most criteria, with formal semantics and modularity. 
Neural networks simulate cognitive processes like memory, learning, and pattern recognition, mimicking brain function \citep{Goodfellow2016}. Neural networks manage complex systems but lack formalization and verification. 
Dynamic systems are used to model continuous behaviors over time, such as emotional dynamics, motivation, and behavioral regulation \citep{Lake2017}. 
Game theory is applied in social psychology to examine how persons make strategic decisions in competitive or cooperative contexts \citep{Sun2016}.
Propositional and modal logic are often used to formalize human reasoning \citep{Marr2015}.
 Perception, attention, and memory are treated as information processing systems, applying entropy and redundancy to explain encoding, storage, and retrieval \citep{Goodfellow2016}.
Markov decision processes (MDP) model decision-making in uncertainty, while agent-based models simulate complex social and ecological interactions \citep{Sun2016}. The last 4 models listed previously (dynamic systems, game theory, propositional and modal logic, and MDPs) largely meet the criteria, offering powerful tools for formal modeling and simulation, though some lack psychological verification or automatic comparability.

  We conclude that \emph{no existing model, tool or method totally satisfy all the 8 previous properties}. More specifically, most models satisfy neither formal verification of properties (property 7) nor decidable comparison between theories (property 8). Surprisingly, almost none of these models use the automaton model, which is the basis of computer science.

\paragraph{Formalization with systems of automata:}
Although the notion of algorithms is used in many fields, few researchers are familiar with the classical models of computability. This is the case in neuroscience and psychology. Yet, the conceptual foundations of computer science would be quite useful for psychology, which also deals with notions of states, actions, behaviors, simulation, and process equivalence, among others. Even computational psychiatry \citep{montague2012computational}, which emerged in the 2010s, does not use computability but rather game theory, probabilistic models, statistics, and machine learning.
Furthermore, finite automata diagrams are intuitive and understandable for researchers without a formal training in mathematics or computer science.

\paragraph{Our contributions:}
\begin{itemize}
\item We present a new automata-based method to formalize psychological theories. In the spirit of \citep{fodor1983modularity}, we build our model by defining and composing modules. Our methodology is based on the principle of modeling different modules with different finite automata which will interact in a very specific way. These modules can be easily modified and refined without changing the whole model. 
\item We provide a method that satisfies the eight properties listed previously.
    \item Our method is demonstrated using the example of stress theory.
 \item We propose a list of new open questions.
\end{itemize}

\vspace{0.2cm}
\noindent
Our (first) modeling, based on finite automata, is only a first stage, and we will continue by adding time (with timed automata), probabilities (with Markov chains), and differential equations on continuous variables.

\paragraph{Structure of the paper:} 
\hyperref[sec:automata]{Section 2} introduces compact-automata and refinements to formalize the development of a theory into successive formal systems. \hyperref[sec:theory]{Section 3} briefly presents the Lazarus and Folkman's (verbal) theory of stress. \hyperref[sec:model]{Section 4} unfolds a step by step refinement process to translate the verbal theory into a formal model: we successively add cognitive appraisal, stress, environment, coping, primary and secondary appraisal and finally commitments.




\section{COMPACT AUTOMATA AND REFINEMENT}
\label{sec:automata}

Throughout this paper, $\Sigma$ will be a non-empty finite set which we call the alphabet. Let us recall that a \emph{finite automaton} (without accepting states) on $\Sigma$ is a tuple $(\Sigma, Q, \delta, I)$, where $Q$ is the finite set of states, $\delta \subseteq Q \times \Sigma \times Q$ is the transition relation, and $I \subseteq Q$ is the finite set of initial states.



\subsection{Synchronized Automata}

There are various systems of (extended) finite automata that synchronise by communication: via rendez-vous \citep{milner_communication_1989,balasubramanian_finding_2023}, blocking or non-blocking \citep{guillou_safety_2023}, broadcast, 
FIFO queues,
or shared memory \citep{atiya1991shared,moiseenko2021survey}.

For the remainder of this article, let $\tau \notin \Sigma$ and $\Sigma_\tau = \Sigma \cup \{ \tau \}$ and $[n] = \{1,2,...,n\}$. Let us now introduce a generalized rendezvous.

\begin{definition}
    
    A \emph{system of $n$ automata synchronized via generalized rendezvous} (short: system)
is a $n$-tuple $\mathcal{S} = (\mathcal{A}_1, \mathcal{A}_2, ..., \mathcal{A}_n) $ where every $\mathcal{A}_i = (\Sigma_\tau, Q_i, \Delta_i, I_i)$ is a finite automaton. 

\end{definition}

The operational semantics of a system $\Sy = (\mathcal{A}_1, \mathcal{A}_2, ..., \mathcal{A}_n)$
is the finite automaton $A(\Sy) =(\Sigma_\tau, Q, \Delta, I)$ defined as follows.
The set of (global) states is $Q = Q_1 \times ... \times Q_n$. For all $q \in Q,$ let us write $q_i$ for the $i^{th}$ element of a (global) state $q = (q_1,..., q_n)$. The set of (global) initial states is $I = I_1 \times ... \times I_n$.
The (global) transition relation, $\Delta \subseteq Q \times \Sigma_\tau \times Q $, is defined by the two following rules: 

\begin{enumerate}
    \item 
\textbf{$\tau$-transitions:} Let $i \in [n]$, $q,q' \in Q$ such that $(q_i, \tau, q_i') \in \Delta_i$, then if for all $ j \in [n] - \{i\}, $ $q_j = q_j'$, there is a $\tau$-transition $q \xrightarrow[]{ \tau} q'$, which stands for $(q,\tau,q') \in \Delta$.

\item
\textbf{Synchronised transitions:} Let $q, q' \in Q, a \in \Sigma$. Let us note $J_a = \{ i \in [n] \mid \exists p_i, r_i \in Q_i$ s.t. $(p_i,a,r_i) \in \Delta_i\}$ the set of indices of the automata that contain at least one $a$-transition. 
If 
%
$J_a \neq \varnothing$, and for all $i \in J_a$, $(q_i, a, q_i') \in \Delta_i$ and 
 for all $i \notin J_a$, $q_i = q_i'$,
%
then there is a synchronised transition $q \xrightarrow[]{a} q'$, which stands for $(q, a, q') \in \Delta$. 
\end{enumerate}
An $a$-transition $(p,a,r) \in \Delta$ 
is enabled when all the automata $\A_i$ such that $i \in J_a$ (i.e., $\A_i$ have at least a local $a$-transition) are ready to enable an $a$-transition. Every $\tau$-transition $(p,\tau,r)$ is local and do not depend on other automata for its execution. There is no blocking condition on $\tau$-transitions.

\begin{example} 

    Consider the system $(\A_1, \A_2, \A_3)$ below:
    
    \begin{figure}[h]
        \begin{center}
    
    \begin{tikzpicture}
    
        \node[state, initial] (q1) {$q_1$};
        \node[state, right of = q1] (q2) {$q_2$};
        
        \draw[->, highlight] (q1) -- (q2) node[midway, above] {$a$};
        \draw[->] (q2)  to[bend right=50] node[midway, above] {$\tau$} (q1);

        \path (q2) edge [loop above] node {$\tau$} (q2);
        
        \node[below= 0.5cm of $(q1)!0.5!(q2)$] {  $\A_1$};
        
        \node[state, right = 1cm of q2] (p) {$p$};
    
        \path (p) edge [loop above, highlight] node {$a$} (p);
        
        \node[below= 0.2cm of p] {  $\A_2$};
        
        \node[state, right = 1cm of p] (r) {$r$};
        
        \path (r) edge [loop above] node {$b$} (r);
        
        \path (r) edge [loop right] node {$\tau$} (r);
        
        \node[below= 0.2cm of r] {  $\A_3$};
       
    \end{tikzpicture}
    
    \end{center}

        \caption{Synchronised transitions and $\tau$-transitions}
        \label{fig:my_label}
    \end{figure}
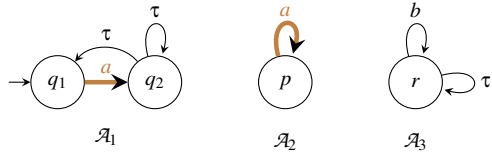

    We have $J_a = \{1,2\}$. Since $3 \notin J_a$ and there are enabled (local) $a$-transitions in (local) states $q_1$ and $p$, we have that $(q_1,p,r) \xrightarrow[]{a} (q_2,p,r)$ is a transition in $\Delta$ (we write $(q_1,p,r) \xrightarrow[]{a} (q_2,p,r) \in \Delta$). 
    
    However, $(q_2,p,r) \xrightarrow[]{a} (q_2,p,r) \notin \Delta$, because $1 \in J_a$ and there are no (local) $a$-transition is enabled in (local) state $q_2$.
    
    Since there is no blocking-condition on $\tau$-transitions, we have that $(q_2, p, r) \xrightarrow[]{\tau} (q_1, p, r) \in \Delta$.
    
    In writing $(q_2,p,r) \xrightarrow[]{\tau} (q_2, p, r) \in \Delta$, it is not precised whether the $\tau-$transition is executed in $\A_1$ or in $\A_3$. This level of precision is enough for what we model. 
    
    Here are two executions of our generalised rendezvous system $A(\A_1,\A_2,\A_3): $

    $(q_1,p,r) \xrightarrow[]{a} (q_2,p,r) \xrightarrow[]{b} (q_2,p,r) \xrightarrow[]{\tau} (q_2,p,r) \xrightarrow[]{\tau} (q_1, p, r) \xrightarrow[]{\tau} (q_1, p, r)$
    
    $(q_1,p,r) \xrightarrow[]{b} (q_1,p,r) \xrightarrow[]{a} (q_2,p,r) \xrightarrow[]{b} (q_2,p,r) \xrightarrow[]{\tau} (q_2,p,r)$

\end{example}

\subsection{Compact-Automata}
\label{sec:compactautomata}

Some of the automata that we will consider in Section~\ref{sec:model} contain a large number of states and transitions. We will thus use \emph{compact-automata} as a way of representing these automata. Compact-automata do not have a behaviour of their own, and they are useful since they are \emph{compact} versions of finite automata, able to represent large and complex automata in a compact and simple way. Below we define compact-automata and the (deterministic) \emph{unfolding} process. Later we will explain the (non-deterministic) \emph{folding} process. Let us fix notations. As previously, $\Sigma_\tau = \Sigma \cup \{\tau\}$ is an alphabet. Let also $\mathcal{V} = \{v_1, v_2, v_3\}$ be a set of three variables, and $\FO$ denotes the set of first order logic formula with relation $=$, variables in $\mathcal{V}$ and with predicate $P_\G$ meaning that $(v_1, v_2, v_3)$ is an edge of $\G$ (for a graph $\G$). 

\begin{definition}
    A \emph{compact-automaton} is a tuple $\tilde{\A} = (\Sigma_\tau, \tilde{Q}, X, f, \tilde{\delta}, I)$, where $\tilde{Q}$ is the set of (compact) states, $X$ is the non-empty set, $f : \tilde{Q} \xrightarrow[]{} \mathcal{P}(X) $ is the unfolding function, $\tilde{\delta} \subseteq \tilde{Q} \times \FO \times \tilde{Q}$ is the (compact) transition relation, and $I \subseteq \cup_{\tilde{q} \in \tilde{Q}} f(\tilde{q})$. 
    
    The unfolding of such $\tilde{\A}$ is $\A = (\Sigma_\tau, Q, \delta, I)$ with $Q = \cup_{\tilde{q} \in \tilde{Q}} f(\tilde{q})$ and where the transition relation $\delta$ of $\A$ is defined by $\delta = \delta_{1} \cup \delta_{2}$, where:
    {\small
    \begin{align}
    & {\small \delta_{1} = \hspace{-2.5em}\bigcup_{(\tilde{q}, \phi, \tilde{p}) \in \tilde{\delta}, \tilde{q} \neq \tilde{p},  a \in \Sigma_\tau}\hspace{-2em} \{ (x, a, y) \mid  x \in f(\tilde{q}), y \in f(\tilde{p}), \phi(x, a, y)\}} \\
    & \delta_{2} =  \bigcup_{(\tilde{q}, \phi, \tilde{q}) \in \tilde{\delta}, a \in \Sigma_\tau} \{ (x, a, x) \mid x \in f(\tilde{q}), \phi(x, a, x) \}
\end{align}}

\end{definition}



In Definition 2, term (1) refers to (compact) transitions where the origin and target (compact) states are different. In this case, the unfolding can generate (unfolded) transitions for all pairs of origin and target (unfolded) states. Term (2) refers to (compact) loops. In this latter case, the unfolding may only generate (unfolded) loops: (with previous notations) if $(\tilde{q}, \phi, \tilde{q}) \in \tilde{\delta}, x,y \in f(\tilde{q}), a \in \Sigma_\tau$ s.t. $x \neq y \land \phi(x, a, y)$, then $(x, a, y)$ is not added to $\delta$ from term (2). 


\begin{figure}[h]
    \begin{center}
        \begin{tikzpicture}

            \node[state] (x) {$\tilde{q}$};
            \node[state, right = 1.5 cm of x] (y) {$\tilde{p}$};
            
            \draw[->] (x) -- (y) node[midway, above] {$(v_2=a)$};
            \path (x) edge [loop below] node {$(v_2=b)$} (x);
            
            \node[below= 1.2cm of $(x)!0.5!(y)$] {compact-automaton $\tilde{\A_1}$};
            \node[draw, rectangle, above= 0.6cm of $(x)!0.5!(y)$] {$I = \{x_0\}, f(\tilde{q}) = f(\tilde{p}) = X$};
            \node[right = 0.2em of y] (arr) {  $\leadsto$};
            
            \node[state, initial, initial text=, right of= arr] (x0) {$x_0$};
            \node[state, right of= x0] (x1) {$x_1$};
            \node[state, right of= x1] (x2) {$x_2$};
        
            \draw[->] (x0) -- (x1) node[midway, above] {$a$};
            \draw[->] (x1) -- (x2) node[midway, above] {$a$};
            \draw[->] (x1)  to[bend right=40] node[midway, above] {$a$} (x0);
            \draw[->] (x2)  to[bend right=40] node[midway, above] {$a$} (x1);
            \draw[->] (x2)  to[bend left=40] node[midway, above] {$a$} (x0);
            \draw[->] (x0)  to[bend left=100] node[midway, above] {$a$} (x2);
        
            \node[below= 1.3cm of $(x1)$] {(finite) automaton $\A_1$ unfolding};
            \node[below= 1.6cm of $(x1)$] {of $\tilde{\A}_1$};
        
            \path (x0) edge [loop below] node {$a,b$} (x0);
            \path (x1) edge [loop above] node {$a,b$} (x1);
            \path (x2) edge [loop below] node {$a,b$} (x2);
        
        \end{tikzpicture}

    \end{center}
    \caption{unfolding of a compact-automaton}
    \label{fig:my_label}
\end{figure}
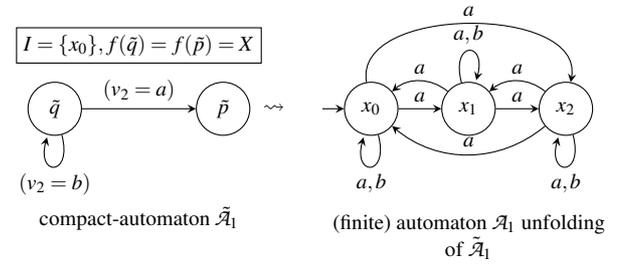

\begin{example} Figure 2 is the unfolding of the compact-automaton $\tilde{\A_1}$ for $X = \{x_0, x_1, x_2\}$.


    
    
    






\end{example}

\paragraph{Graphical representation of compact-automata.} Let $\tilde{\A} = (\Sigma_\tau, \tilde{Q}, X, f, \tilde{\delta}, I)$ be a compact-automaton. We use three types of graphical notations in compact-automata diagrams:
\begin{enumerate}
    \item \textbf{Sets and singletons:}\label{par:compactsets} Let $\tilde{q} \in \tilde{Q}$. If $\mid f(\tilde{q}) \mid > 1$, we represent the (compact) state $\tilde{q}$ as a double circle around $\tilde{q}$. If $f(\tilde{q}) = \{ q \}$, we represent the (compact) state $\tilde{q}$ as a simple circle around $\tilde{q}$ (same representation as for finite automata). 
    
    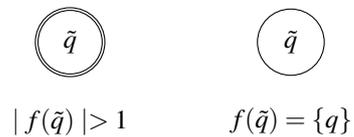
\begin{figure}[h]
        \begin{center}

        \begin{tikzpicture}[>=stealth, node distance=2cm, scale=1, every node/.style={scale=1}]
        
            \node[state, accepting] (x) {$\tilde{q}$};
            \node[below = 0.3cm of x] {$\mid f(\tilde{q}) \mid > 1$};

            \node[state, right = 2cm of x] (xj) {$\tilde{q}$};
            \node[below = 0.3cm of xj] {$f(\tilde{q}) = \{ q \}$};
        
        \end{tikzpicture}
        \end{center}
        
        \caption{graphical representations of sets and singletons}
        \label{fig:my_label}
    \end{figure}

     \item \textbf{Transitions with a letter or a variable:} when an element of $\FO$ is of the form $(v_2 = a)$ for a letter $a \in \Sigma_\tau$, we will simply label the transition as $a$. When an element of $\FO$ is of the form $[v_1] = v_2$, and the origin state is written $[\tilde{q}]$, we can simply label the transition with $\tilde{q}$. 

     \item \textbf{Graph transitions:} \label{par:graphtransition} if $(\tilde{q}, \phi, \tilde{p}) \in \tilde{\delta}$, and there is a directed labeled graph $\G$, with labels of $\G$ belonging to $L \subseteq \Sigma_\tau$, with $f(\tilde{q}) \neq f(\tilde{p})$, $f(\tilde{q}) = f(\tilde{p}) = X$, and the vertices of $\G$ are $X$. If $\phi(v_1,v_2,v_3) = ((v_1, v_2, v_3)$ is an edge of $\G)$, then we label the (compact) transition $\G/L$.
     
     \begin{figure}[h]
          \begin{tikzpicture}
            
                \node[state, accepting] (x) {$x$};
                \node[state, accepting, right = 1cm of x] (y) {$y$};
                
                \draw[->] (x) -- (y) node[midway, above] {$\G / L$};

                \node[above= 1.5cm of $(x)!0.5!(y)$] {Representation for};
                \node[above= 1cm of $(x)!0.5!(y)$] {$\phi(v_1,v_2,v_3) = ((v_1, v_2, v_3)$ is an edge of $\G)$};
    
                \node[below= 0.5cm of $(x)!0.5!(y)$] {compact-automaton $\tilde{\A}_2$};
                
                \node[right = 0.3cm of y] (plus) {$+$};
                \node[fill=black, circle, inner sep=0.5pt, right = 0.3cm of plus] (x0) {$x_0$};
                \node[above= 0.2cm of $(x0)$] {$x_0$};
                \node[below= 1cm of x, xshift = -1cm] (=) {$\leadsto$};
                \node[fill=black, circle, inner sep=0.5pt, right of= x0] (x1) {$x_1$};
                \node[above= 0.2cm of $(x1)$] {$x_1$};
                \node[fill=black, circle, inner sep=0.5pt, above right of= x0] (x2) {$x_2$};
                \node[above= 0.2cm of $(x2)$] {$x_2$};
            
                \draw[->] (x0) -- (x1) node[midway, above] {$a$};
                \draw[->] (x0) -- (x2) node[midway, above] {$b$};
    
                \path (x1) edge [loop below] node {$b$} (x1);
    
                \node[state, below = 2cm of x] (x0') {$x_0$};
                
                \node[state, right of= x0'] (x1') {$x_1$};
                \node[state, above right of= x0'] (x2') {$x_2$};
                
                \node[below= 0.5cm of $(x0')!0.5!(x1')$] {Unfolding of $\tilde{\A}_2$ for $X = \{x_0, x_1, x_2\}$, };
                \node[below= 1cm of $(x0')!0.5!(x1')$] {$\G$ and $L = \{a, b\}$};

                \draw[->] (x0') -- (x1') node[midway, above] {$a$};
                \draw[->] (x0') -- (x2') node[midway, above] {$b$};
                \path (x1') edge [loop right] node {$b$} (x1');
    
                \node[below= 0.5cm of $(x0)!0.5!(x1)$] {$\G$};

            \end{tikzpicture}
         \caption{graphical representation of graph transitions}
         \label{fig:my_label}
     \end{figure}
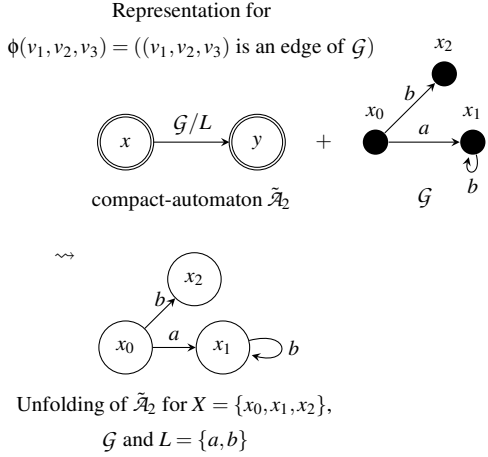


        
            
        
            
        


 
 \end{enumerate}
 
 \paragraph{Folding.} The folding process is non-deterministic. From a large finite automaton, first we identify sets of states for which the structure of transitions is similar; second, we create a small number of (compact) states and (compact) transitions that represent all the states and transitions of the initial automaton.

\subsection{Refinements of Automata}
\label{sec:refinement}

The notion of refinement helps us understand how successive systems relate to one another: a more refined system is modeled more precisely with regards to the theory. Refinement between systems is expressed as a binary relation. In order to define it, we used two intermediary definitions which correspond to refinements over labels and transitions, and over automata. 

Let us consider two systems $\Sy = (\mathcal{A}_1, \mathcal{A}_2, ..., \mathcal{A}_n)$ and $\Sy' = (\mathcal{A}_1', \mathcal{A}_2', ..., \mathcal{A}_m')$ with their associated automata $A(\Sy) = (\Sigma_\tau, Q, \Delta, I)$ and $A(\Sy') = (\Sigma_\tau', Q', \Delta', I')$. As usual, we note $\Sigma_\tau = \Sigma \cup \{ \tau \}$ and $\Sigma_\tau' = \Sigma' \cup \{ \tau \}$.

\begin{definition} \textbf{Refinement of labels and transitions:}
    We define the following relation $\leq_1$ on $\Sigma_\tau \cup \Sigma_\tau'$ as the smallest relation satisfying the following condition:
    
    For all labels $\lambda \in \Sigma_\tau \cup \Sigma_\tau'$, $ \tau \leq_1 \lambda$ and $\lambda \leq_1 \lambda$.

    We extend this relation to elements of $\Delta \cup \Delta'$ in the following way : for all $q_1, q_2 \in Q, p_1,p_2, \in Q', \lambda \in \Sigma_\tau, \lambda' \in \Sigma_\tau'$ we have $(q_1, \lambda, p_1) \leq_1 (q_2, \lambda', p_2)$ if and only if $\lambda \leq_1 \lambda'$.
\end{definition}

\begin{definition}\textbf{Refinement of automata:} 
let us consider $i \in [n], j \in [m]$. We say that $\A_i = (\Sigma_\tau, Q_i, \Delta_i, I_i) \leq_2 \A_j' = (\Sigma_\tau', Q_j', \Delta_j', I_j')$ iff there exists a partition $(Q_q)_{q \in Q_i}$ of $Q_j'$ for which the following conditions hold:
\begin{itemize}
    \item $ \Sigma_\tau \subseteq \Sigma_\tau'$
\end{itemize}

\begin{itemize}
     \item \emph{Transition refinement}: for all $t'=(p', \lambda', q') \in \Delta'_j$, for $p, q \in Q_i$ such that $p' \in Q_{p}$ and $q' \in Q_{q}$, there exists a transition $t= (p, \lambda, q)$ such that $ t \leq_1 t'$
    \item \emph{Refinement of initial states}: for all $q' \in I_j'$, $\exists q \in I_i, q' \in Q_q$
\end{itemize}

\end{definition}

\begin{example}
Figure 6 explicits all possible ways of refining transitions. Since there are no loops on $q_2$ there cannot be any transition between $p_1$, $p_2$, and $p_3$. 

\begin{figure}[h]
        \begin{center}
            
        \begin{tikzpicture}[node distance = 1.4cm]
        
        \node[state,initial, fill = blue!20]    (A)                     {$q_1$};
        \node[state, fill = brown!20]  (B) [right = 3cm of A]        {$q_2$};
        \node[left= 0.1cm of A, yshift = 0.5cm] {Automaton $\A_i$};
        
        \path
        (A) edge[bend left, ->]  node[midway, above] {$\tau$} (B)
        (B) edge[bend left, ->]  node[midway, above] {$a$} (A);
        \path (A) edge [loop above] node {$ \tau$} (A);
        
        \node[state,initial, fill = blue!20]    (X1) [below=of A]        {$p_1$};
        \node[state, fill = blue!20]            (X2) [right of=X1]       {$p_2$};
        \node[state, fill = brown!20]            (X3) [right = 0.5cm of X2]       {$p_3$};
        \node[state, fill = brown!20]            (X4) [right of=X3]       {$p_4$};
        \node[state, fill = brown!20]  (X5) [right of=X4]       {$p_5$};
        
         \node[left= 0.1cm of X1, yshift = -0.5cm] {Automaton $\A_j'$}; 
         \node[left= 0.1cm of X1, yshift = -1cm] {refinement of $\A_i$};
        
        \path
        (X1) edge[bend left, ->] node[midway, above] {$\tau$} (X2)
        (X2) edge[bend left, ->] node[midway, above] {$b$} (X4)
        (X5) edge[bend left=40, ->] node[midway, above] {$a$} (X1);
        
        \path (X2) edge [loop above] node {$c$} (X2);
        
        \draw[<->, dashed, blue] (A) to[out=-135, in=135] (X1);
        \draw[<->, dashed, blue] (A) to[out=-135, in=135]  (X2);
        \draw[<->, dashed, brown] (B) to[out=-45, in=45]  (X3);
        \draw[<->, dashed, brown] (B) to[out=-45, in=45] (X4);
        \draw[<->, dashed, brown] (B) to[out=-45, in=45]  (X5);
        
        \end{tikzpicture}
        
        \end{center}

    \caption{Refinement of an automaton}
    \label{fig:my_label}
\end{figure}
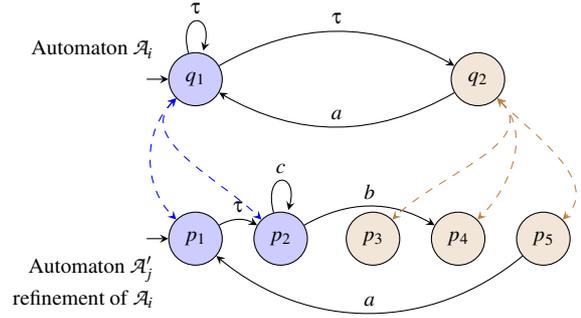

\end{example}

Let us recall that for all $a \in \Sigma_\tau$, $J_a = \{ i \in [n] \mid \exists p_i, r_i \in Q_i$ s.t. $(p_i,a,r_i) \in \Delta_i\}$ is the set of indices of the automata that contain at least one $a$-transition. For system $\Sy'$ and $a \in \Sigma'$ we note it $J_a'$.

\begin{definition}
    \textbf{Refinement of systems: } For $\Sy = (\mathcal{A}_1, \mathcal{A}_2, ..., \mathcal{A}_n)$ and $\Sy' = (\mathcal{A}_1', \mathcal{A}_2', ..., \mathcal{A}_m')$, we say that $\Sy \leq_3 \Sy'$ if and only if: 
    \begin{itemize}
        \item $n \leq m$
        \item For all $1 \leq i \leq n$, $\A_i \leq_2 \A_i'$
        \item For all $a \in \Sigma, J_a \subseteq J_a'$
    \end{itemize}
\end{definition}

\begin{remark}
    The three relations $\leq_1, \leq_2,$ and $\leq_3$ are quasi-orders. 
\end{remark}

\section{LAZARUS AND FOLKMAN'S THEORY OF STRESS}
\label{sec:theory}

This section explains the elements of Lazarus and Folkman's theory of stress that we have identified as both important and suitable for modeling, see section~\ref{sec:model}. These elements come from the book \textit{Stress, appraisal and coping} \citep{lazarus_stress_nodate}, and hence, every reference to pages in this section points to pages of it. This book\footnote{Which has gathered more than 95000 quotes according to google scholar as of now} is arguably one of the most representative of this theory. We detail these elements below, namely the relational definition of stress, cognitive appraisal, primary appraisal, secondary appraisal, coping, emotional-focused coping, problem-focused coping, and commitments. 

\paragraph{Relational meaning of stress.} \label{par:stress} Until the 1960s, stress was defined either as a stimulus (stressor) (p.12), or as a physiological response (stress response) (p.14). These definitions fail to explain the discrepancies in response between persons to the same stimulus (p.19). This motivated Lazarus and Folkman to introduce a \emph{relational} definition of stress, taking both aspects into account: "Psychological stress is a particular relationship between the person and the environment that is appraised by the person as taxing or exceeding his or her resources and endangering his or her well-being" (p.19, l.26). 

\paragraph{Cognitive appraisal} \label{par:appraisal}is the process by which a person determines whether a relationship is stressful or not (p.19). This process is continuous during waking life (p.31) and contains two main processes: \emph{primary appraisal} and \emph{secondary appraisal}(p.31), which "cannot be considered as separate" (p.43, l.8). 

\paragraph{Primary appraisal} \label{par:pa} is the process through which a person answers the question of whether a particular stimulus is beneficial, irrelevant or stressful (p.32).

\paragraph{Secondary appraisal} \label{par:sa}is the process through which a person evaluates the coping strategies available to deal with a stimulus appraised as stressful (p.35). 


\paragraph{Coping} \label{par:coping}is defined as "constantly changing cognitive and behavioral efforts to manage specific external and/or internal demands that are appraised as taxing or exceeding the resources of the person" (p.141, l.10). Furthermore, it is said that the dynamic of coping are "a function of the continuous appraisals (...) of the evolving person-environment relationship" (p.142, l.32). 

\paragraph{Emotion-focused coping and problem-focused coping.} \label{par:efc} Coping strategies fall into two categories: emotion-focused coping efforts are meant to regulate emotions, whereas problem-focused coping efforts are meant to resolve the situation itself (p.150).

\paragraph{Commitments} \label{par:commitments}"express what is important to the person, what has meaning for him or her."(p.56, l.4) They play a key role in the appraisal process, since the stress experienced by a person is directly related to how much meaning is tied to its commitments. 




\section{A MODEL FOR LAZARUS AND FOLKMAN'S THEORY OF STRESS}
\label{sec:model}

This section unfolds a step by step refinement process to translate a verbal theory into a formal model. This methodology consists in starting from a trivial initial system, and at each step, either adding a new concept from the theory, or refining a previously modeled concept. At each step, new concepts or developments must be based on previously modeled concepts. From this condition, next concepts or developments to model are chosen with regards to their importance and their simplicity. Every new system is a refinement of the previous one. In this paper, the whole methodology is not formally defined. 

In the remainder of this paper, we will refer to compact-automata as automata, referring to the unfolding of a compact-automaton when it is mentioned. Subsequent systems are noted $\Sy_1, \Sy_2, \Sy_3,...$, and for all natural number $i$, and (compact or normal) automata of system $\Sy_i$ will be noted $\A_{i.1},\A_{i.2},\A_{i.3},...$ 

All explanations from Section~\ref{sec:theory} were based on the book \textit{Stress, appraisal and coping} \citep{lazarus_stress_nodate}. Likewise, we will base our modeling solely on this book, and hence, every reference to pages in this section points to pages of the book.

\subsection{The Initial System}

We begin with a (trivial) initial system $\Sy_1 = (\A_{1.1})$, which is meant to represent a system where nothing yet has been specified. The universe is represented as having an unique state, with some observed activity. When we model a dynamic taking place which we do not specify, we add a $\tau$-transition. This is the role of the $\tau$-loop on state $uni$. 
\begin{figure}[h]
    \begin{center}
    \begin{tikzpicture}[>=stealth, node distance=4cm, scale=1, every node/.style={scale=1}]
        \node[state, initial] (r) {uni};
        
        \path (r) edge [loop above] node {$\tau $} (r);
    
        \node[below = 0.2cm of r] {$\A_{1.1} :$ initial automaton};
    \end{tikzpicture}

    \end{center}
    
    \caption{system $\Sy_1$}
    \label{fig:my_label}
\end{figure}

$\mathcal{A}_{1.1} = (\Sigma_{\tau, 1}, Q_{1.1}, \Delta_{1.1}, I_{1.1})$ with $\Sigma_{\tau, 1} = \{ \tau \}$, $Q_{1.1} = \{ uni \}$, $\Delta_{1.1} = \{(\text{uni}, \tau, \text{uni})\}$ and $I_{1.1} = \{\text{uni} \}$. 
 
\subsection{Adding Cognitive Appraisal}

Here we model the key notion of \hyperref[par:appraisal]{cognitive appraisal}\footnote{blue-colored text points to other sections of the paper} defined as follows:
\emph{"Cognitive appraisal can be most readily understood as the process of categorizing an encounter, and its various facets, with respect to its significance for well-being(...) it is largely evaluative, focused on meaning or significance, and takes place continuously during waking life" (p.31, l.18)}



Since what happens outside of waking life does not seem to be explicited in the book, for the sake of simplicity, we decide that appraisal happens \emph{only} during waking life. Since we know from the theory that some activity takes place during appraisal, we add a $\tau$-loop on the "appraisal" state. The fact that the loop is labeled by $\tau$ means that we do not specify the nature of this activity. Nothing is mentioned in the theory about activity during the "non-awake" state, and hence we do not add a $\tau$-loop on top of this state. 
This leads to the following system $\Sy_2 = (\A_{2.1})$ with one automaton with two states.

\begin{figure}[h]
    \begin{center}

    \vspace{-0.2cm}
    \begin{tikzpicture}[>=stealth, node distance=4cm, scale=1, every node/.style={scale=1}]

        \node at (2, -2.4) {$\A_{2.1}$ : Cognitive appraisal};
        \node[state, initial] (r) {non-awake};
        \node[state] [right of = r](a) {appraisal};
    
        \draw[->] (r) -- (a) node[midway, above] {$\tau$};
        \draw[->] (a)  to[bend left=40] node[midway, above] {$\tau$} (r);

        \path (a) edge [loop above] node {$ \tau$} (a);

    \end{tikzpicture}

\end{center}
    \caption{system $\Sy_2$}
    \label{fig:my_label}
\end{figure}

Formally, the system $\Sy_2$ consists only of automaton $\A_{2.1}$, where ${A}_{2.1} = (\Sigma_{\tau, 2}, Q_{2.1}, \Delta_{2.1}, I_{2.1})$ is defined by $\Sigma_{\tau} = \{ \tau \}$, $Q_{2.1} = \{ \text{non-awake}, \text{awake} \}$, $\Delta_{2.1} = \{(\text{non-awake}, \tau, \text{appraisal}),$\\$ (\text{appraisal}, \tau, \text{appraisal}),(\text{appraisal},\tau, \text{non-awake})\}$, and $I_{2.1} = \{\text{non-awake} \}$.

System $\Sy_2 = (\A_{2.1})$ is a \hyperref[sec:refinement]{refinement} of system $\Sy_1 = (\A_{1.1})$: $ 1 \leq 1$, $\A_{1.1} \leq \A_{2.1}$,  and since $\Sigma_{\tau, 1} = \{\tau\}$, the third condition is always true. $\A_{1.1} \leq \A_{2.1}$ because: if $Q_{uni} = \{ non-awake, appraisal \}$, then $(Q_{uni})$ is clearly a partition of $Q_{2.1}$. For all transitions $(q, \tau, p)$ of $\A_{2.1}$ the transition $(uni, \tau, uni)$ satisfies $\tau \leq \tau$ and $p,q \in Q_{uni}$. Finally, $I_{2.1} = \{ \text{non-awake} \}$ and $\text{non-awake} \in Q_{uni}$.

\subsection{Adding Stress}
\label{sec:stress}



In Section~\ref{par:stress} we discussed the \hyperref[par:stress]{relational definition of stress}. For now we will only distinguish between two consequences of cognitive appraisal : "no-stress", and "stress". 

\vspace{0.4cm}

To our previous system $\Sy_2$, we add a second automaton that calculates stress. The calculation of stress can only happen during the appraisal state of the first automaton. This way of modeling these sentences is not unique, but will be coherent with the further development of the theory.

Automata $A_{3.1}$ and $\A_{3.2}$ synchronise on two letters: "stress" and "no-stress". The set of automata that possess a (local) $stress$-transition is $J_{stress} = \{(3.1),(3.2)\}$. Hence, both automata must have an enabled (local) $stress$-transition for a global $stress$-transition to take place. This is why in (global) state $(non-awake, f)$, a (global) $stress$-transition cannot occur, whereas in (global) state $(appraisal, f)$, the (global) $stress$-transition is enabled. Here is an execution of the system:

$(non-awake, f) \xrightarrow[]{\tau} (appraisal, f) \xrightarrow[]{stress} (appraisal, f) \xrightarrow[]{no-stress}  (appraisal,f) \xrightarrow[]{\tau} $ \\ $  (non-awake, f)$

This execution described a person who wakes up, appraises the person-environment relationship as stressful, appraises again this relationship as non-stressful, then goes back to sleep. 

\begin{figure}[h]
    \begin{center}

\begin{tikzpicture}
    
    \node[state, initial] (r) {non-awake};
    \node[state, right = 0.5cm of r] (a) {appraisal};
    
    \node[below = 2.1cm of $(r)!0.5!(a)$] {$\A_{3.1}$ : Cognitive appraisal};

    \draw[-> ] (r) -- (a) node[midway, above] {$\tau$};
    \draw[-> ] (a) to[bend left=40] node[midway, above] {$\tau$} (r);
    \path (a) edge[loop above, highlight] node {$stress$} (a);
    \path (a) edge[loop below, highlight] node {$no-stress$} (a);
    \path (a) edge[loop right ] node {$\tau$} (a);

    \node at (current bounding box.south) [below=2cm, right = 2cm] {};

    \node[state, initial, highlight, right = 1.8cm of a] (f) {$f$};
    
    \node[below = 1.5cm of f] {$\A_{3.2}$ : Calculation of stress};
    
    \path (f) edge[loop above, highlight] node {$stress$} (f);
    \path (f) edge[loop below, highlight] node {$no-stress$} (f);
    \path (f) edge[loop right, highlight] node {$\tau$} (f);
    
\end{tikzpicture}

\end{center}
    \caption{system $\Sy_3$}
    \label{fig:my_label}
\end{figure}

In the remainder of this article, we will note $na$ for "non-awake", $a$ for "appraisal", $s$ for "stress" and $\overline{s}$ for "no-stress". Formally, $\Sy_3 = (\A_{3.1}, \A_{3.2})$, $\Sigma_\tau = \{s, \overline{s}, \tau\}$, and:

\begin{tikzpicture}
    
    \def\rectWidth{0.22\textwidth}
    
    \node[draw, anchor=north west, minimum width=\rectWidth, minimum height=2cm] (rect1) at (0,0)
    {\parbox{0.5\textwidth}{
    $\mathcal{A}_{3.2} = (\Sigma_{\tau,3}, Q_{3.2}, \Delta_{3.2}, I_{3.2}), Q_{3.2} = \{ f \} $ \\ $\Delta_{3.2} = \{ (f, s, f), (f, \overline{s}, f), (f, \tau, f) \} $ \\ $ I_{3.2} = \{f\}.$
   }};

    \def\rectWidth{0.22\textwidth}

    \node[draw, anchor= west, minimum width=\rectWidth, minimum height=2cm] (rect2) at (0, 1)
    {\parbox{0.5\textwidth}{ $\mathcal{A}_{3.1} = (\Sigma_{\tau,3.1}, Q_{3.1}, \Delta_{3.1}, I_{3.1}), Q_{3.1} = \{ na, a \}$ \\ $\Delta_{3.1} = \{(na, \tau, a), (a, s, a), (a, \overline{s}, a), (a, \tau, na), (a, \tau, a)\}$ \\ $I_{3.1} = \{r \}.
    $
    
    }};
\end{tikzpicture}

System $\Sy_3 = (\A_{3.1}, \A_{3.2})$ is a \hyperref[sec:refinement]{refinement} of system $\Sy_2 = (\A_{2.1})$: $ 1 \leq 2$, $\A_{2.1} \leq \A_{3.1}$, and since $\Sigma_{\tau,2} = \{ \tau\}$, the third condition is always true. $\A_{2.1} \leq \A_{3.1}$ because if $Q_{\text{non-awake}} = \{ na \}$ and $Q_{\text{appraisal}} = \{ a\}$ then $(Q_{\text{non-awake}}, Q_{\text{appraisal}})$ is a partition of $Q_{3.1} = \{na, a\}$. The two added transitions in $\Delta_{3.2}$ from $\Delta_{3.2}$ are $(a, s, a)$ and $(a, \overline{s}, a)$, the other ones remaining exactly the same. $(\text{appraisal}, \tau, \text{appraisal})$ satisfies $\tau \leq s$, $\tau \leq \overline{s}$, and $a \in Q_\text{appraisal}$. Finally, initial states are not modified. In the remainder of this paper, we will not detail the proof that each new system is a refinement of the previous one.


\subsection{Adding the Environment}

We consider that the environment is a \hyperref[sec:compactautomata]{compact-automaton} $\A_{4.3} = (\Sigma_\tau, \tilde{Q}_3, X, f, \tilde{\delta}_3, I_3)$, on parameters $(X, \G_\tau)$, where $X$ is a finite set of states and $\mathcal{G}_\tau$ is a directed labeled graph whose vertices are $X$ and labels are $\tau$. Since this is the first compact-automaton of our modeling, we detail it here. In the remainder of this paper, the details will only appear on the graphical representations. We have $\tilde{Q} = \{x, y\}$, and $f(x) = f(y) = X$. We have that $\tilde{\delta} = \{ (x, (v_1 = [v_2]), y), (x, ((v_1, v_2, v_3) \text{ is an edge of } \G_\tau), y) \}$. The second transition of $\tilde{\delta}$ is graphically labeled by $\G_\tau / \{ \tau \}$, see Subsection \hyperref[par:graphtransition]{2.2}.



Appraisal happens "continuously during waking life"(p.31, l.23) \hyperref[par:appraisal]{(see Section 3)}. One way of modeling the continuity of the appraisal process during waking life is to have the "Environment" automaton "send" periodic updates of its state to the "Cognitive appraisal" automaton while it is in the "appraisal" state. 

This is done via synchronised transitions: automata $\A_{4.1}$ and $\A_{4.3}$ synchronise over letters of $X$. On automaton $\A_{4.1}$, the transition labeled by $X$ means that this transition exists for all $x \in X$. 
Let's fix $x \in X$. The name of the state associated with $x$ in the unfolded automaton of $\A_{4.3}$ is $[x]$, to avoid confusion between states and transitions of the automaton. Since  $J_x = \{ (4.1), (4.3)\}$, the (global) $x$-transition $(a,f,[x]) \xrightarrow[]{x} (a,f,[x])$ is enabled. Automaton $\A_{4.3}$ in state $[x]$ synchronises on letter $x$ with $\A_{4.1}$ when it is in state $a$. 

Here is an execution, when we have $x_1 \in X$ such that $(x_0, \tau, x_1)$ is an edge of $\mathcal{G}_\tau$:

$(na, f, [x_0]) \xrightarrow[]{\tau} (a,f,[x_0]) \xrightarrow[]{x_0} (a,f,[x_0]) \xrightarrow[]{s} (a,f,[x_0]) \xrightarrow[]{\tau} (a,f,[x_1]) \xrightarrow[]{x_1} (a,f,[x_1]) \xrightarrow[]{\overline{s}} (a, f, [x_1])$

This execution represents a person who wakes up, perceives its environment, appraises the person-environment relationship as stressful, the environment changes by itself, the person perceives it, and appraises the new person-environment relationship as non-stressful. 

\begin{figure}[h]
    \begin{center}
\begin{tikzpicture}

    \node[state, initial] (r) {$na$};
    \node[state, right of=r] (a) {$a$};

    \node[below = 1.5cm of $(r)!0.5!(a)$]{  $\A_{4.1}$ : Cognitive appraisal};

    \draw[-> ] (r) -- (a) node[midway, above] {$\tau$};
    \draw[-> ] (a) to[bend left=40] node[midway, above] {$\tau$} (r);
    \path (a) edge[loop above ] node {$s, \overline{s}$} (a);
    \path (a) edge[loop below, highlight] node {$X$} (a);
    \path (a) edge[loop right] node {$\tau$} (a);

    \node[state, initial, right = 1.5cm of a] (f) {$f$};
    \node[below = 1.2cm of $(f)$, xshift = 1cm] {  $\A_{4.2}$ : Calculation of stress};
    
    \path (f) edge[loop above ] node {$s$} (f);
    \path (f) edge[loop below ] node {$\overline{s}$} (f);
    \path (f) edge[loop right ] node {$\tau$} (f);

    \node[state, accepting, highlight, below= 3cm of r] (x) {$[x]$};
    \node[state, right = 2cm of x, highlight, accepting] (y) {$[y]$};
    \node[below = 0.5cm of $(x)!0.5!(y)$] {  $\A_{4.3}$ : Environment};
    \node[above = 0.8cm of $(y)$, draw, rectangle] {$I_{4.3} = \{[x_0]\}, f([x]) = f([y]) = X$};

    \draw[->, highlight] (x) -- (y) node[midway, above] {$\G_\tau / \{\tau\}$} ;

    \path (x) edge[loop above, highlight] node {$x$} (x);


\end{tikzpicture}

\end{center}
    \caption{system $\Sy_4$}
    \label{fig:my_label}
\end{figure}

\subsection{Adding Coping}

We introduce a "Coping"\hyperref[par:coping]{(see explanations in Section 3)} automaton that models the decision of engaging in coping efforts. Once a "stress" appraisal has been calculated in $\A_{5.2}$ as a (global) $s$-transition, the "Coping" automaton gets to a state from which it can "instruct" the "Environment" automaton to engage with coping efforts, via new synchronised transitions. 

\label{par:graph}As the environment, the modeling of coping has parameters $(C, \mathcal{G}_C)$, where $C$ is a finite set of coping strategies, and $\mathcal{G}_C$ is a directed labeled graph whose vertices are $X$ and whose edges are labeled by elements of $C$. The "Coping" automaton $\A_{5.4}$ has two states: $\rho$ for "rest" and $[C]$ for the state where coping efforts can be engaged. The "Environment" automaton was modified by adding transitions corresponding to the edges of $\mathcal{G}_C$: for all $x,z \in X, c \in C$ such that $(x,c,z)$ is an edge of $\mathcal{G}_C$, $([x],c,[z])$ is a transition of automaton $\A_{5.3}$. Each of these (local) transitions can synchronise with automaton $\A_{5.4}$ when it is in state $[C]$. Although it is clear that humans can act when they are not stressed, we only consider here coping efforts which occur during psychological stress. 

Let's consider $x_1 \in X, c \in C$ such that $(x_0, c, x_1)$ is an edge of $\mathcal{G}_C$. Here is a possible execution of our new system:

$(na, f, [x_0], \rho) \xrightarrow[]{\tau} (a, f, [x_0], \rho) \xrightarrow[]{x_0} (a, f, [x_0], \rho) \xrightarrow[]{s} (a, f, [x_0], [C]) \xrightarrow[]{c} (a, f, [x_1], \rho)  \xrightarrow[]{x_1} (a, f, [x_1], \rho) \xrightarrow[]{\overline{s}} (a, f, [x_1], \rho)$

This execution represents a person who wakes up, perceives its environment, appraises the person-environment relationship as stressful, engages in coping efforts with coping strategy $c$, perceives the changed person-environment relationship, appraises it as non-stressful.

\begin{figure}[h]
    \begin{center}

\begin{tikzpicture}

    \node[state, initial] (r) {$na$};
    \node[state, right = 1.5cm of r] (pa) {$a$};

    \node[below = 1.3cm of $(r)!0.5!(a)$] {$\A_{5.1}$ : Cognitive appraisal};

    \draw[-> ] (r) -- (pa) node[midway, above] {$\tau$};
    \draw[-> ] (pa) to[bend left=40] node[midway, above] {$\tau$} (r);

    \path (pa) edge[loop above ] node {$\overline{s}, s$} (pa);
    \path (pa) edge[loop below] node {$X$} (pa);
    \path (pa) edge[loop right] node {$\tau$} (pa);

    \node[state, below=3cm of r, accepting] (x) {$[x]$};
    \node[state, right = 1.5cm of x, accepting] (y) {$[y]$};
    \node[below = 0.5cm of $(x)!0.5!(y)$] {$\A_{5.3} :$ Environment};

    \draw[->] (x) -- (y) node[midway, above] {$\G_\tau / \tau$};
    
    \draw[->, highlight] (x) to[bend left=40] node[midway, above] {$\G_C / C$} (y);
    
    \path (x) edge[loop above] node {$x$} (x);
    
    \node[above = 1.1cm of $(y)$, draw, rectangle, xshift = -1cm] {$I_{5.3} = \{[x_0]\}, f([x]) = f([y]) = X$};


    \node[state, initial, right= 1.5cm of pa] (f) {$f$};
    \node[below = 1cm of f] {$\A_{5.2}$ : Calculation of stress};
    
    \path (f) edge[loop above ] node {$s$} (f);
    \path (f) edge[loop below ] node {$\overline{s}$} (f);
    \path (f) edge[loop right ] node {$\tau$} (f);
    
    \node[state, initial, below=3cm of f, highlight, xshift= -0.5cm] (s') {$\rho$};
    \node[state, highlight] [right = 1.5cm of s'](s) {$[C]$};
    \node[below = 1cm of $(s')!0.5!(s)$] {$\A_{5.4}$ : Coping};

    \draw[->, highlight] (s') -- (s) node[midway, above] {$s$};
    \draw[->, highlight] (s)  to[bend left=40] node[midway, above] {$C$} (s');

    \path (s') edge [loop above, highlight] node {$\overline{s}$} (s');
    
\end{tikzpicture}
\end{center}
 
    \caption{system $\Sy_5$}
    \label{fig:my_label}
\end{figure}

The notation $C$ on a transition means that this transition exists for any letter $c \in C$.


\subsection{Refining Appraisal into Primary Appraisal and Secondary Appraisal}

\hyperref[par:pa]{Primary} and \hyperref[par:sa]{secondary} appraisal are the two main processes of cognitive appraisal. 

To model these two processes, we refine the state $a$ of automaton $\A_{5.1}$ into two states $pa$ and $sa$. The function of \hyperref[par:pa]{primary appraisal} is the one conducted by automaton $\A_{5.2}$, which remains unchanged as automaton $\A_{6.2}$ which we rename "Primary appraisal". Since secondary appraisal happens once a stimulus is appraised as \hyperref[par:sa]{stressful}, and that coping efforts are a function of \hyperref[par:coping]{previous appraisals}, it seems fitting to include secondary appraisal into the "Coping" automaton. The system loops between states $pa$ and $sa$ for a while as a way of determining \emph{"the degree of stress and the quality (or content) of the emotional reaction"} (p.35, l.29). 

Secondary appraisal happens after "stress" has been calculated by automaton $\A_{6.2}$ ((global) $s$-transition) and before engaging in coping efforts. We refine automaton $\A_{5.4}$ by refining state $[C]$ into $2|C|$ states: $[c]$ and $[[c]]$ for all $c \in C$. We note $C' = \{ [c], c \in C\}$ and $C'' = \{ [[c]], c \in C\}$. A (global) $s$-transition leads non-deterministically to one state $[c] \in C'$ which will be evaluated. Two results are possible, coming as two synchronised transitions between $\A_{6.1}$ and $\A_{6.4}$: ”good” (letter $g$), or ”bad” (letter $b$). For a $c \in C$, a "good" evaluation from state $[c]$ brings $\A_{6.1}$ back to state $pa$, and $\A_{6.4}$ to state $[[c]] \in C''$, from which coping efforts are engaged for coping strategy $c$. A "bad" evaluation from state $[c]$ will lead $\A_{6.1}$ back to state $pa$ and $\A_{6.4}$ back to state $\rho$. From this global state, another primary appraisal can occur, and another strategy can be evaluated. This is the loop between the states $pa$ and $sa$ mentioned earlier. 


Let us describe how the model $\Sy_6=(\A_{6.1}, \A_{6.2}, \A_{6.3},\A_{6.4})$ with $c_1, c_2 \in C$, $x_1 \in X$ such that $(x_0, c_2, x_1)$ is an edge of $\mathcal{G}_C$, \emph{formalizes} the following human sequence: \emph{a person wakes up, perceives its environment, appraises the person-environment relationship as stressful, wonders if coping strategy $c_1$ would be beneficial, perceives $c_1$ as a bad strategy, wonders if coping strategy $c_2$ would be beneficial, perceives $c_2$ to be a good strategy, engages in coping efforts with strategy $c_2$, the environment changes, the person perceives the new environment, the person appraises the new person-environment relationship as non-stressful.} 

Here is the corresponding execution :

$(na, f, [x_0], \rho) \xrightarrow[]{\tau} (pa, f, [x_0], \rho) \xrightarrow[]{x_0} (pa, f, [x_0], \rho) \xrightarrow[]{s} (sa, f, [x_0], [c_1]) \xrightarrow[]{b} (pa, f, [x_0], \rho) \xrightarrow[]{s} (sa, f, [x_0], [c_2]) \xrightarrow[]{g} (sa, f, [x_0], [[c_2]]) \xrightarrow[]{c_2} (pa, f, [x_1], \rho) \xrightarrow[]{x_1} (pa, f, [x_1], \rho) \xrightarrow[]{\overline{s}} (pa, f, [x_1], \rho)$. 

\begin{figure}[h]
    \begin{center}

\begin{tikzpicture}

    \node[state, initial] (r) {$na$};
    \node[state, right = 1cm of r, highlight] (pa) {$pa$};
    \node[state, right = 1cm of pa, highlight] (sa) {$sa$};

    \node[below = 1cm of pa] {  $\A_{6.1}$ : Cognitive appraisal};

    \draw[-> ] (r) -- (pa) node[midway, above] {$\tau$};
    \draw[-> ] (pa) to[bend left=40] node[midway, above] {$\tau$} (r);
    \draw[->] (pa) -- (sa) node[midway, above] {$s$};
    \draw[->, highlight] (sa) to[bend left=40] node[midway, above] {$g,b$} (pa);
    \path (pa) edge[loop above ] node {$\overline{s}$} (pa);
    \path (pa) edge[loop below] node {$X$} (pa);

     \node[state, below=3cm of r, accepting] (x) {$[x]$};
    \node[state, right = 1.5cm of x, accepting] (y) {$[z]$};
    \node[above = 1.1cm of $(y)$, draw, rectangle, xshift= -1cm] {$I_{6.3} = \{[x_0]\}, f([x]) = f([y]) = X$};
    \node[below = 0.7cm of $(x)!0.5!(y)$] {  $\A_{6.3} :$ Environment};

    \draw[->] (x) -- (y) node[midway, above] {$\G_\tau / \{\tau\}$};
    
    \draw[->] (x) to[bend left=40] node[midway, above] {$\G_X / C$} (y);
    
    \path (x) edge[loop above] node {$x$} (x);

    \node[state, initial, right = 1cm of sa] (f) {$f$};
    \node[below = 1.5cm of $(f)$] {  $\A_{6.2}$ : Calculation of stress};
    
    \path (f) edge[loop above ] node {$s$} (f);
    \path (f) edge[loop below ] node {$\overline{s}$} (f);
    \path (f) edge[loop right ] node {$\tau$} (f);
    
    \node[state, initial, right =0.7cm of y, highlight] (rho) {$\rho$};
    \node[state, accepting, highlight] [right = 1cm of rho](c) {$[c]$};
    \node[state, accepting, below=1cm of c, highlight] (c') {$[[c]]$};

    \draw[->, highlight] (rho) -- (c) node[midway, above] {$s $};
    \draw[->, highlight] (c)  to[bend left=40] node[midway, above] {$b$} (rho);
    \draw[->, highlight] (c')  to[bend left=40] node[midway, above] {$c$} (rho);
    \draw[->, highlight] (c) -- (c') node[midway, right, yshift=0.2cm] {$([v_1] = v_3) $} node[midway, right, yshift=-0.2cm] {$\land (v_2 = g)$};

    \path (rho) edge[loop above, highlight] node {$\overline{s}$} (rho);
    
    \node[above = 0.5cm of c, draw, rectangle, align=center] 
    {$f([c]) = C',f([[c]]) = C''$};
    \node[below = 2cm of c] {  $\A_{6.4}$ : Secondary appraisal}; 
    \node[below=2.3cm of c]{and coping};

\end{tikzpicture}
\end{center}
    \caption{system $\Sy_6$}
    \label{fig:my_label}
\end{figure}
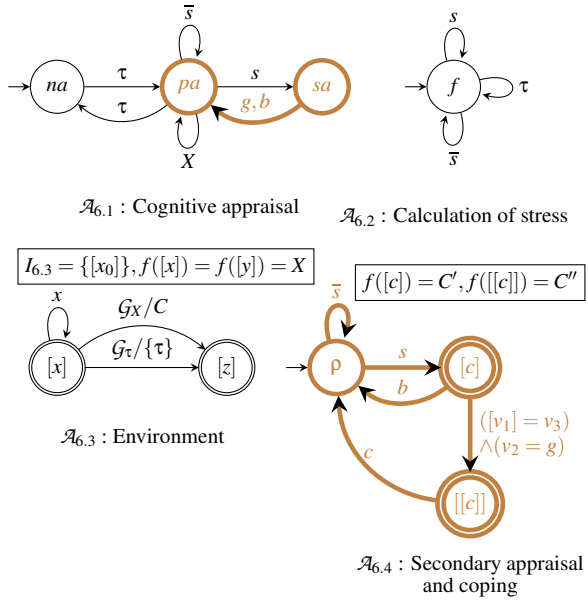

\begin{remark}
    To our knowledge, the decision-making mechanism between secondary appraisal and coping is not detailed in \citep{lazarus_stress_nodate}. We have chosen to model it in a very simple way: coping efforts are engaged when a coping strategy has been evaluated as "good". Let's note that any decision making theory could be incorporated here instead. 
\end{remark}

\subsection{Adding Commitments}

As seen in \hyperref[par:commitments]{Section 3}, commitments play an important role in the appraisal process. 


We model a commitment via a functions $\varphi : X \xrightarrow[]{} \{0,1\}$. The set of all such functions is noted $\Phi$. At this point, the set of internal variables is $\Phi$ and the set of global states of the person-environment whole is $X \times \Phi$.

We interpret the functions $\varphi$ as follows: in global state $(x,\varphi)$, the individual is stressed if and only if $\varphi(x) = 0$.








Since $\Phi$ is finite, we add a new "Internal parameters" automaton $\A_{6.5}$. Transitions between states of this automaton are coping efforts intending to affect internal parameters \hyperref[par:efc]{(Emotion-focused forms of coping, see Section 3)}. The graph $\mathcal{G}_C$ introduced in \hyperref[par:graph]{Subsection 4.5} is now renamed $\mathcal{G}_X$. The modeling of coping affecting internal parameters has parameters $(\Phi, \mathcal{G}_\Phi),$ where $\mathcal{G}_\Phi$ is a directed labeled graph whose vertices are $\Phi$ and whose edges are labeled by elements of $C$. For all $c \in C$, the (global) $c$-transition is extended to automaton $\A_{6.5}$ which then synchronises with $\A_{6.3}$ and $\A_{6.4}$. 

\begin{figure}[h]
    \begin{center}

    \begin{tikzpicture}

    \node[state, initial] (r) {$na$};
    \node[state, right = 1cm of r] (pa) {$pa$};
    \node[state, right = 1cm of pa] (sa) {$sa$};

    \node[below = 1cm of pa] {  $\A_{7.1}$ : Cognitive appraisal};

    \draw[-> ] (r) -- (pa) node[midway, above] {$\tau$};
    \draw[-> ] (pa) to[bend left=40] node[midway, above] {$\tau$} (r);
    \draw[->] (pa) -- (sa) node[midway, above] {$s$};
    \draw[->] (sa) to[bend left=40] node[midway, above] {$g,b$} (pa);
    \path (pa) edge[loop above ] node {$\overline{s}$} (pa);
    \path (pa) edge[loop below] node {$X$} (pa);

     \node[state, below=3cm of r, accepting] (x) {$[x]$};
    \node[state, right = 1.5cm of x, accepting] (y) {$[z]$};
    \node[above = 1.1cm of $(y)$, draw, rectangle, xshift= -1cm] {$I_{7.3} = \{[x_0]\}, f([x]) = f([y]) = X$};
    \node[below = 0.7cm of $(x)!0.5!(y)$] {  $\A_{7.3} :$ Environment};

    \draw[->] (x) -- (y) node[midway, above] {$\G_\tau / \{\tau\}$};
    
    \draw[->] (x) to[bend left=40] node[midway, above] {$\G_X / C$} (y);
    
    \path (x) edge[loop above] node {$x$} (x);

    \node[state, initial, right = 1cm of sa] (f) {$f$};
    \node[below = 1.5cm of $(f)$] {  $\A_{7.2}$ : Calculation of stress};
    
    \path (f) edge[loop above ] node {$s$} (f);
    \path (f) edge[loop below ] node {$\overline{s}$} (f);
    \path (f) edge[loop right ] node {$\tau$} (f);
    
    \node[state, initial, right =0.7cm of y] (rho) {$\rho$};
    \node[state, accepting] [right = 1cm of rho](c) {$[c]$};
    \node[state, accepting, below=1cm of c] (c') {$[[c]]$};

    \draw[->] (rho) -- (c) node[midway, above] {$s $};
    \draw[->] (c)  to[bend left=40] node[midway, above] {$b$} (rho);
    \draw[->] (c')  to[bend left=40] node[midway, above] {$c$} (rho);
    \draw[->] (c) -- (c') node[midway, right, yshift=0.2cm] {$([v_1] = v_3) $} node[midway, right, yshift=-0.2cm] {$\land (v_2 = g)$};

    \path (rho) edge[loop above] node {$\overline{s}$} (rho);
    
    \node[above = 0.5cm of c, draw, rectangle] {$f([c]) = C', f([[c]]) = C''$};

    \node[below = 2cm of c] {  $\A_{7.4}$ : Secondary appraisal}; 
    \node[below=2.3cm of c]{and coping};

    \node[state, accepting, below=2cm of x, highlight] (phi) {$\varphi$};
    \node[state, accepting, highlight] [right =2cm of phi](phi') {$\varphi'$};

    \draw[->, highlight] (phi) -- (phi') node[midway, above] {$\G_\Phi / C$};
    
    \node[above = 0.6cm of $(phi)!0.5!(phi')$, draw, rectangle] {$I_{7.5} = \{\varphi_0\}, f(\varphi) = f(\varphi') = \Phi$};

    \node[below = 0.5cm of $(phi)!0.5!(phi')$] {  $\A_{7.5}$ : Internal parameters};

\end{tikzpicture}
\end{center}
  
    \caption{system $\Sy_7$}
    \label{fig:my_label}
\end{figure}
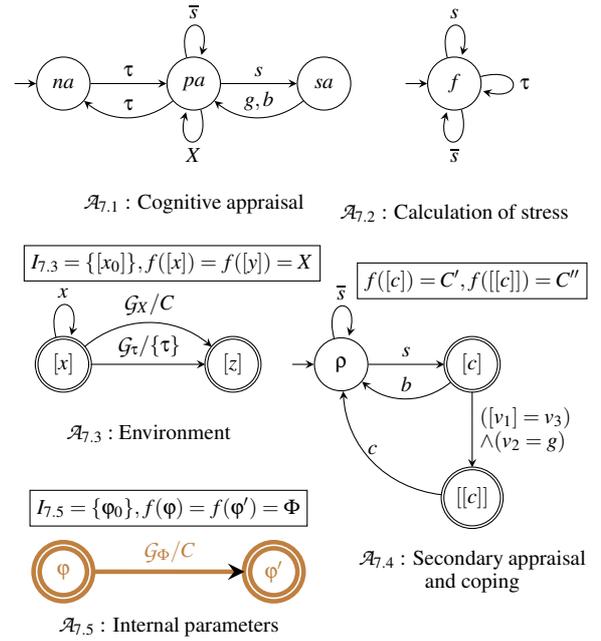

Here is a simple example to show how this formalism can be implemented. 

\begin{example} Here is a person's description of his relationship with money:

"It's important for me to have enough money. I want to feel like I'm safe financially. Enough money for me is having more than 1000 euros in my bank account. Right now I have enough money. Sometimes people steal money from my bank account and I have no money left. As a way to make myself feel better, I try to save a lot of money each month, and I try to think that money is not so important"

There are two states of the world 

$X = \{\geq 1000, <1000\} $, where $\geq 1000$ (resp. $<1000$) is the state where there is more (resp. strictly less) than $1000$ euros in the person's bank account.

Let's define $\varphi: X \xrightarrow[]{}\{0, 1\}$ such that $\varphi(\geq 1000) = 1$ and $\varphi(<1000) = 0$ which corresponds to the commitment expressed by the person. 

$\Phi = \{\varphi, \mathbf{1} - \varphi, \mathbf{1}, \mathbf{0}\}$, where $\mathbf{1}$ (resp. $\mathbf{0}$) is the function always equal to $1$ (resp. to $0$).

The only $\tau$-transitions in the environmnent automaton $\A_{6.3}$ correspond to "Sometimes people steal money from my bank account and I have no money left". Hence the two edges of $\mathcal{G}_\tau$ are $(\geq 1000, \tau, <1000)$ and $(<1000, \tau, <1000)$. 

$C = \{ c_1, c_2 \}$

The coping strategy $c_1$ corresponds to "saving money". This strategy only affects the environment. The coping strategy $c_2$ corresponds to "trying to think money is not so important". This strategy only affects internal parameters. Hence, the edges of $\mathcal{G}_X$ are : 

$(<1000, c_1, <1000), (<1000, c_1, \geq 1000), $\\$ (\geq 1000, c_1, \geq 1000), (<1000, c_2, <1000), $\\$ (\geq 1000, c_2, \geq 1000)$. 

The edges of $\mathcal{G}_\Phi$ are: 

$(\varphi, c_2, \varphi), (\varphi, c_2, \mathbf{1}), (\mathbf{1}, c_2, \mathbf{1}), (\varphi, c_1, \varphi), (\mathbf{1}, c_1, \mathbf{1})$. 

A transition $(\varphi, c_2, \mathbf{1})$ corresponds to a shift in the person's commitment with money. 

With this example, here are the diagrams of automata $\A_{7.3}$ and $\A_{7.5}$:

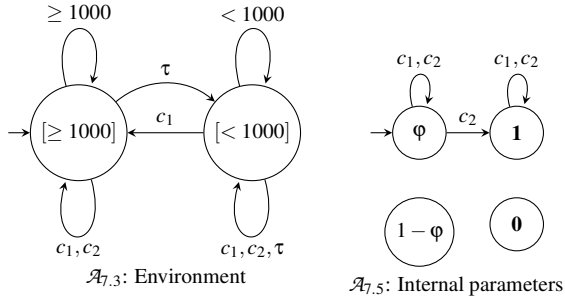
\begin{figure}[h]
\begin{center}

\begin{tikzpicture}

    \node[state, initial] (x1) {$[\geq 1000]$};
    \node[state, right = 1cm of x1] (x2) {$[<1000]$};

    \draw[->] (x2) -- (x1) node[midway, above] {$c_1$};
    \draw[->] (x1) to[bend left=40] node[midway, above] {$\tau$} (x2);

    \path (x1) edge[loop above] node {$\geq 1000$} (x1);
    \path (x1) edge[loop below] node {$c_1, c_2$} (x1);
    \path (x2) edge[loop below] node {$c_1, c_2, \tau$} (x2);
    \path (x2) edge[loop above] node {$<1000$} (x2);

    \node[below = 1.7cm of $(x1)!0.5!(x2)$] {  $\A_{7.3}$: Environment};

    \node[state, initial, right =1.2cm of x2] (y1) {$\varphi$};
    \node[state, right of = y1] (y2) {$\mathbf{1}$};
    \node[state, below =0.5cm of y1] (y3) {$1 - \varphi$};
    \node[state, below =0.5cm of y2] (y4) {$\mathbf{0}$};
    
    \path (y1) edge[loop above] node {$c_1, c_2$} (y1);
    \path (y2) edge[loop above] node {$c_1, c_2$} (y2);
    
    \draw[->] (y1) -- (y2) node[midway, above] {$c_2$};

    \node[below = 1.8cm of $(y1)!0.5!(y2)$, xshift= -0.2cm] {$\A_{7.5}$: Internal parameters};
    
\end{tikzpicture}

\end{center}

    \caption{Example of modeled environment and internal parameters}
    \label{fig:my_label}
\end{figure}

\end{example}

\subsection{Further Developments}

It is possible to refine the concept of \hyperref[par:pa]{primary appraisal} by calculating a result ("stress" or "non-stress") based on the state of $\A_{7.3} \times \A_{7.5}$.

The calculation of \hyperref[par:sa]{secondary appraisal} can also be modeled by adding an automaton which corresponds to the inner representations of the person-environment relationship, with an imagination process. 

It is also possible to add other results of primary appraisal, namely beneficial, irrelevant, harm/loss, threat or challenge (p.32-33). For this, we can simply add other letters instead of $s$ and $\overline{s}$. 

Several concepts or mechanism can be modelled by adding internal parameters to automaton $\A_{7.5}$ and by changing the commitment functions to take the new internal parameters into account. This can be done for the interaction between primary and secondary appraisal (p.35), for beliefs about personal control(p.65), as well as the concept of novelty (p.83).

\section{\uppercase{Conclusion and perspectives}}
\label{sec:conclusion}



\textbf{Conclusion:}
We presented a new automata-based method to formalize psychological theories, with an example applied to stress theory. We demonstrated how to create increasingly precise systems. Our modelization allows for a large number of modules that is not possible with a verbal theory.\\
    
\noindent
\textbf{Perspectives:}
We open the way to many directions of research:
\begin{itemize}
\item Since we have represented psychological concepts by both states and transitions, it is natural to define the \emph{language of a theory $\mathcal{T}$ modelized by a system $\Sy_n$} either as the set of accepted words in $\Sigma_{\tau}^*$ (all states are final) of the system $\Sy_n$ or as the set of accepted sequences of transitions in $\Delta^*$ (where $\Delta \subseteq Q \times \Sigma_{\tau} \times Q $). We now have a canonical formal object, i.e., a finite automaton, allowing us to leverage the full range of conceptual and practical tools from theoretical computer science. We are now able to compare our theory with other theories also expressed by automata.
\item Let $\Sy_{i+1}$ be a system obtained by refinement of system $\Sy_i$; prove (under realistic hypothesis) some formal properties of refinement like $L(\Sy_{i+1}) \subseteq L(\Sy_{i})$ or $\Sy_i$ simulates $\Sy_{i+1}$ (for a simulation ordering to choose).
\item We plan to apply our methodology and to build formal models of many other theories like cognitive theory of emotions, properties of short-term memory, long-term memory, perception, representations of the world, imagination, rational thinking,...
\item We will try to formalize into the computability framework two theories of the mind and consciousness like the Friston theory and the GWT.
  \item We will also extend our class of models by adding probabilities, time and continuous variables.
  \item We will implement and automatically verify some psychological theories with existing tools like state-chart \citep{harel_statecharts_1987}, event-b \citep{schneider_behavioural_2014}, NuSMV \citep{cimatti2002nusmv} (A symbolic model checker), SPIN \citep{holzmann1997model} (A tool for the formal verification of distributed software systems) and UPPAAL \citep{DBLP:conf/sfm/BehrmannDL04} (A tool for modeling, simulation, and verification of real-time systems) \citep{holzmann1997model}.
  \item We will host seminars and workshops to train researchers in psychology to model theories with this tool. This will be the opportunity to validate the usefulness and usability of our tool.
\end{itemize}

\section*{\uppercase{Acknowledgements}}

We thank the reviewers of the Modelsward conference for their insightful comments.

\bibliographystyle{apalike}
{\small
\bibliography{example}}

\end{document}